\documentclass[conference]{IEEEtran}
\IEEEoverridecommandlockouts
\usepackage{cite}
\usepackage{amsmath,amssymb,amsfonts}
\usepackage{algorithmic}
\usepackage{graphicx}
\usepackage{textcomp}
\usepackage{xcolor}
\usepackage[normalem]{ulem}
\def\BibTeX{{\rm B\kern-.05em{\sc i\kern-.025em b}\kern-.08em
    T\kern-.1667em\lower.7ex\hbox{E}\kern-.125emX}}
\begin{document}

\title{Impact of Terminal Noise on Polarization Rotation Vector for Sensing Applications\\
\thanks{This work has been performed in the framework
of the ECSTATIC project
with GA 101189595. Miquel Masanas performed the lab experiment and akcnowledges SENSEI with GA 101189545. Both received funding from the European Union’s Horizon Europe Framework Programme.}
}

\author{\IEEEauthorblockN{1\textsuperscript{st} Mohammad M. Hosseini}
\IEEEauthorblockA{\textit{Hardware \& Engineering} \\
\textit{Nokia}\\
Munich, Germany \\
mohammad.hosseini@nokia.com}\
\and
\IEEEauthorblockN{2\textsuperscript{nd}Miquel Masanas}
\IEEEauthorblockA{\textit{Hardware \& Engineering} \\
\textit{Nokia}\\
Munich, Germany \\
miquel.1.masanas@nokia.com}
\and
\IEEEauthorblockN{3\textsuperscript{rd} Giuseppe Parisi}
\IEEEauthorblockA{\textit{Hardware \& Engineering} \\
\textit{Nokia}\\
Munich, Germany \\
giuseppe.parisi@nokia.com}
\and
\IEEEauthorblockN{4\textsuperscript{th} Antonio Mecozzi}
\IEEEauthorblockA{\textit{Physical and Chemical Sciences} \\
\textit{Universit{\`a} dell'Aquila}\\
L'Aquila, Italy \\
antonio.mecozzi@univaq.it}
\and
\IEEEauthorblockN{5\textsuperscript{th} Alberto Marullo}
\IEEEauthorblockA{\textit{Operations - Submarine Specialist} \\
\textit{Sparkle}\\
Rome, Italy \\
alberto.marullo@tisparkle.com}
\and
\IEEEauthorblockN{6\textsuperscript{th} Danilo Decaroli}
\IEEEauthorblockA{\textit{Head of Operations} \\
\textit{Sparkle}\\
Rome, Italy \\
danilo.decaroli@tisparkle.com}
\and
\IEEEauthorblockN{7\textsuperscript{th} Antonio Napoli}
\IEEEauthorblockA{\textit{Hardware \& Engineering} \\
\textit{Nokia}\\
Munich, Germany \\
antonio.napoli@nokia.com}
}

\maketitle
\begin{abstract}
State-of-Polarization sensing with coherent transponders enables wide-area geophysical monitoring over existing submarine cables, but its performance is limited by polarization noise from both the fiber and terminal hardware. This work investigates how terminal noise affects polarization rotation estimates derived from receiver equalizer coefficients and how it can obscure stochastic polarization drift used for sensing. We analyze Jones-matrix time series from two deployed receivers in the Sparkle Mediterranean link MedNautilus (approximately 2000 km and 450 km) and compare them with a laboratory back-to-back reference. Power Spectral Density (PSD) analysis reveals a low-frequency random-walk regime and a high-frequency white-noise floor, separated by a link-dependent corner frequency. The rotation innovation variance increases with link length, while the longest field link also shows elevated white noise consistent with accumulated amplifier and terminal contributions. Additionally, harmonic spectral components are observed, indicating a transponder-related artifact that should be considered in practical sensing applications.
\end{abstract}


\section{Introduction}
Global submarine optical networks, which form the invisible, transoceanic backbone of the modern digital economy, offer vast and, historically, underutilized potential for cost-effective, planetary-scale environmental monitoring. Originally engineered exclusively for high-capacity data transmission across ocean basins, these telecommunication cables are increasingly recognized as an extensive, ready-made sensory apparatus. By leveraging State-of-Polarization (SOP) sensing, existing coherent transponders can detect dynamic seismic and oceanic perturbations over intercontinental distances exceeding $9000\text{ km}$~\cite{mecozzi2026earth}. Because SOP monitoring fundamentally requires environmental stability -- to isolate subtle transient anomalies from background chaotic fluctuations -- it is highly effective in the low-noise, isothermal, and mechanically insulated deep-sea conditions that characterize the abyssal plains of the ocean floor~\cite{mecozzi2025geophysical, zhan2021optical}. %Because SOP monitoring requires environmental stability, it is highly effective in low-noise deep-sea conditions~\cite{mecozzi2025geophysical}. 
The primary sensing mechanism relies on the sophisticated, high-speed adaptive Multiple-Input Multiple-Output (MIMO) equalizers embedded within modern digital coherent receivers. These Digital Signal Processing (DSP) units are designed to track and compensate for transmission-induced linear impairments, including polarization effects, in real time~\cite{sun2020800g}. In performing this continuous compensatory function, the DSP algorithms can extract the optical link’s complete input-output Jones matrix directly from the MIMO filters coefficients. However, the estimation of the Jones matrix can be inaccurate due to different sources of noises involved in the process~\cite{savory2010digital}. Also, unwanted perturbations, such as temperature changes acting on the cable and the transponder itself, mechanical vibrations, and most critically, the inherent polarization rotational random walk noise—driven by the fiber's static random birefringence—injects a chaotic, non-deterministic drift into the SOP that complicates the target signal detection~\cite{czegledi2016polarization}. 

\begin{figure}[htb]
    \centering
    \includegraphics[width=0.95\linewidth]{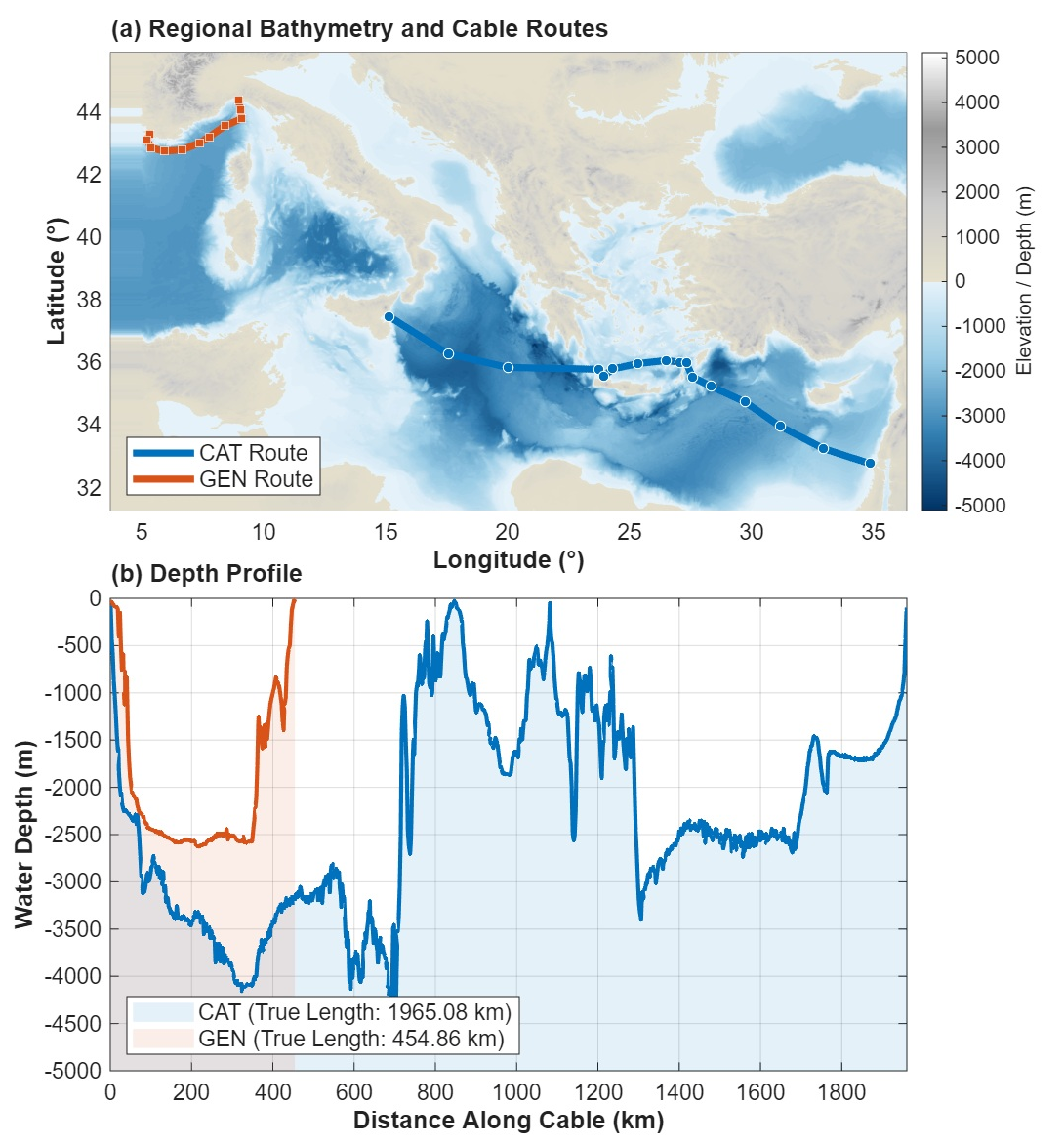}
    \caption{(a) Cable routes of $\sim$2000 km (CAT) and $\sim$450 km (GEN) and (b) depth profile of the routes using bathymetry data from the GEBCO~\cite{gebco2026}.}
    \label{fig:map}
\end{figure}

Traditionally, several works have been carried out to characterize polarization dynamic induced by optical fibers deployed in different environments in the context of optical communication using polarimeters and coherent transponders~\cite{gordon2000pmd,hauske2009optical}. These investigations have focused on quantifying Polarization Mode Dispersion (PMD) fluctuations, Differential Group Delay (DGD) drift, and the evolution of the output SOP over various timescales~\cite{karlsson2000long}. Historical measurements, dating back to the 1980s, primarily used polarimeters to monitor installed submarine and terrestrial cables, observing that polarization drift is typically slow (on the order of hours to days) but can become rapid in response to mechanical vibrations, wind, or significant temperature changes. More recently, the development of digital coherent transponders has enabled monitoring of these dynamics by extracting channel parameters directly from transponders~\cite{geyer2008channel,mazur2022transoceanic}.

In this study, we analyze multiple data sets acquired from the same transponders deployed in the subsea Sparkle network in the Mediterranean Sea and in a laboratory environment~\cite{sun2020800g}. Our objective is to investigate polarization from a sensing perspective and to determine the performance limits and capabilities of sensing using coherent transponders. We first review the theoretical modeling of ideal polarization dynamics, then discuss the practical challenges, and finally present a data analysis to support and validate the proposed concepts.

\begin{figure*}[h!]
    \centering
    \includegraphics[width=0.95\linewidth, trim={2.7cm 0.7cm 2.7cm 1.2cm},
        clip]{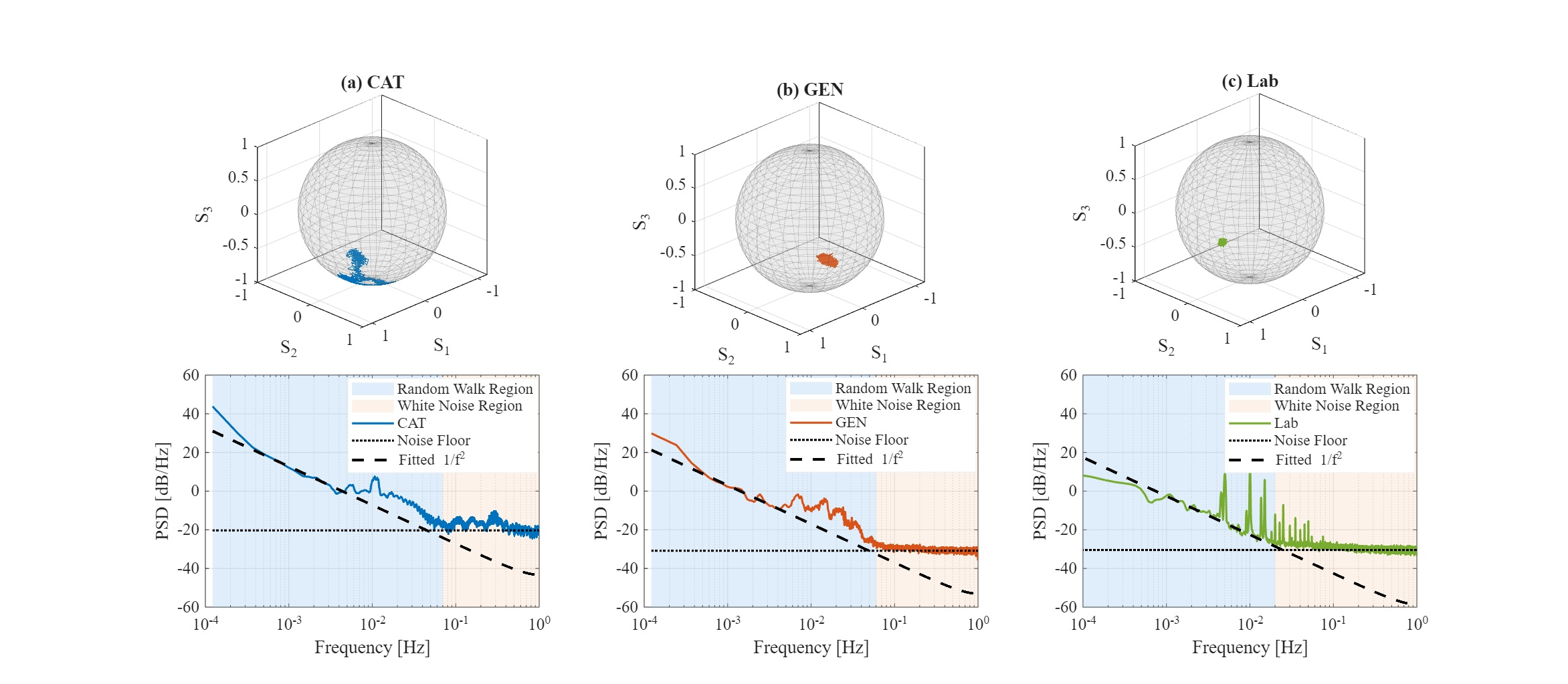}
    % \put(-245,195){(a)}
    % \put(-125,195){(b)}
    % \put(-245,95){(c)}
    % \put(-125,95){(d)}
    \caption{Stokes representation and Jones rotation vector total PSD for (a) CAT, (b) GEN and (c) Lab transponders.}
    \label{fig:psd}
\end{figure*}

\section{Polarization Dynamic}\label{sectionII}
Modeling optical polarization dynamics is essential for evaluating telecommunication performance, particularly for understanding PMD and SOP drift. There are two primary mathematical spaces and several physical frameworks to characterize these effects. In the Jones formalism, the optical field is represented by $2 \times 1$ vector while the transformation is represented by a complex ($2 \times 2$) unitary matrix, known as the Jones matrix. This representation is particularly well suited for modeling cascaded fiber segments, since the overall channel response can be obtained through sequential matrix multiplication. Because the Jones formalism is defined in the (SU(2)) space whereas polarization rotations are observed in the (SO(3)) Stokes space, the mapping between the two representations is two-to-one. Consequently, the Jones matrices ($\mathbf{J}$) and (-$\mathbf{J}$) correspond to the same rotation of the Stokes vector on the Poincaré sphere, leading to an intrinsic sign ambiguity when reconstructing Jones matrices from measurements~\cite{karlsson2000long}. In contrast, the Stokes formalism represents the SOP through observable optical power quantities, mapping the SOP onto a three-dimensional real-valued vector on the Poincaré sphere. This approach is especially advantageous for visualization and for statistical modeling of polarization evolution, where the dynamics can be interpreted as a stochastic random walk on the sphere. In this formalism the transformations are described by Muller matrices~\cite{czegledi2016polarization}. The waveplate model represents a single-mode fiber as a chain of many short birefringent sections, each with its own orientation, length, and refractive properties, with the overall transmission behavior obtained by multiplying their Jones matrices. Building on this, the hinge model of PMD reflects real fiber networks, where stable buried fiber sections are connected by short exposed segments (``hinges'') that fluctuate due to environmental effects~\cite{antonelli2006theoretical}. The isotropic hinge model treats these hinges as completely random polarization rotators, while the anisotropic (waveplate) hinge model more realistically assumes rotations around fixed axes with varying angles, leading to different predictions for outage probability and channel performance.

In the model proposed in~\cite{czegledi2016polarization}, the time evolution of the polarization is captured by modeling the Jones matrix as a sequence of random matrices. This approach effectively emulates a random walk on the Poincar'e sphere. Unlike previous approaches in the literature, that typically treated the Jones matrix as constant or governed by a deterministic cyclic or quasi-cyclic rotation pattern, this theoretical framework generalizes the one-dimensional phase noise random walk to rotational random walk with multiplicative innovation instead of additive. By simply adding a matrix, the resulting matrix is no longer unitary. Physically, this means the model is now suddenly inventing or absorbing optical power out of nowhere. The model operates by updating the channel matrix over time using a succession of random innovation matrices, where each matrix corresponds to a time increment and utilizes all three physical degrees of freedom to describe the temporal evolution of the SOP. Mathematically, the time evolution of the Jones matrix $J_k$ at the $k$-th time instance is formulated as $J_k = J(\vec{\dot{\phi}}_k) J_{k-1}$. Here, $J(\vec{\dot{\phi}}_k)$ denotes the random innovation matrix, which is defined via the exponential matrix as $J(\vec{\dot{\phi}}_k) = \exp(-i \frac{\vec{\dot{\phi}}_k}{2} \cdot \vec{\sigma})$, where $\vec{\sigma}$ represents the Pauli vector. This innovation matrix is directly parameterized by the random vector $\vec{\dot{\phi}}_k$, which is independently drawn at each time step from a zero-mean Gaussian distribution such that $\frac{\vec{\dot{\phi}}_k}{2} \sim \mathcal{N}(0, \sigma_p^2 I_3)$, with $I_3$ being the $3 \times 3$ identity matrix. The variance of this distribution is set to $\sigma_p^2 = 2\pi \Delta p T$. Within this variance expression, $T$ is the symbol interval, and $\Delta p$ defines the polarization linewidth parameter, which strictly quantifies the speed of the stochastic SOP drift analogous to how laser linewidth describes phase noise.
% \begin{table}[ht]
%     \centering
%     \caption{Rotation vector variance before and after drift compensation/filtering}
%     \label{tab:dataset_metrics}
%     % 'S' columns automatically format the numbers provided in the cells
%     \small
%     \begin{tabular}{l S S}
%         \toprule
%         \textbf{Dataset} & {\textbf{Uncompensated}} & {\textbf{Compensated}} \\
%         \midrule
%         CAT & 9.8561    & 0.00460 \\
%         GEN & 0.20096   & 0.00295 \\
%         Lab & 0.0034625 & 0.00072 \\
%         \bottomrule
%     \end{tabular}
%     \normalsize
% \end{table}
\section{Results}
We collected the Jones matrix data from operational coherent transponders deployed in two cable routes of Sparkle networks in Mediterranean sea as shown in Fig.~\ref{fig:map} as well as the 1D depth profile approximated with the help of bathymetric data obtained from the GEBCO global grid~\cite{gebco2026}. The lab transponder is in back-to-back mode with a received OSNR of 38 dB at 800 Gbps in probabilistic constellation shaping format. All data are stored at a 2 Hz sampling rate without any anti-aliasing filtering before acquisition. For a pure $\frac{1}{f^2}$ process, decimation is usually safe and the folded noise from higher bands should be negligible. However, when there is flat white noise, this has to be carefully assessed. The CAT cable is around 2000 km long, while the cable of the GEN transponder is around 450 km.

\begin{table}[h!]
\centering
\caption{Frequency and noise variance parameters}
\begin{tabular}{|l|c|c|c|}
\hline
\textbf{} & \textbf{Corner Freq} & $\mathbf{\sigma_p^2}$\text{[$rad^2$]} & \textbf{White Noise VAR \text{[$rad^2$]}} \\ \hline
\textbf{CAT} & $\sim 70 \text{ mHz}$ & $1.6 \times 10^{-5}$ & $9.3 \times 10^{-3}$ \\ \hline
\textbf{GEN} & $\sim 60 \text{ mHz}$ & $1.6 \times 10^{-6}$ & $8.3 \times 10^{-4}$ \\ \hline
\textbf{LAB} & $\sim 20 \text{ mHz}$ & $4.5 \times 10^{-7}$ & $9.0 \times 10^{-4}$ \\ \hline
\end{tabular}

\label{tab:noise_params}
\end{table}

First, we computed the Power Spectral Density (PSD) of the Jones rotation vector obtained from 24 hours of continuous measurements, using the preprocessing method proposed by A. Mecozzi~\textit{et al.}~\cite{mecozzi2025geophysical0}. The resulting PSD exhibits mixed behavior, characterized by a low frequency regime where random-walk is dominant and a white-noise floor at higher frequencies.

To quantify this behavior, we identified the corner frequency as the intersection point between the extrapolated random-walk and white-noise regions for each case. Based on this separation, we estimated $\sigma_p^2$ from the low-frequency random-walk regime and the white-noise variance from the high-frequency flat region. Note that without this separation, the random walk speed can be overestimated. The resulting parameters are summarized in Table~\ref{tab:noise_params}. The estimated $\sigma_p^2$ shows a clear correlation with the cable length, suggesting that optical components within the transponders and the transmission fibers contribute significantly to these variations. In particular, the CAT transponder exhibits approximately 10 times and 35 times larger rotation-angle innovation variance compared to the GEN, and Lab transponders respectively. This indicates that, although the CAT route is only slightly more than four times longer than the GEN route, it exhibits more than four times higher noise levels. This increase can be attributed to the larger number of terminal points along the CAT route, where the cable comes to the surface and undergoes more environmental noise. In contrast, the white-noise contribution remains at a similar level for the GEN and Lab transponders, with a variance of approximately $8.5 \times 10^{-4}$. However, the CAT link exhibits a white-noise level that is nearly one order of magnitude higher. This noise may originate from Amplified Spontaneous Emission (ASE) noise~\cite{pellegrini2025overview}, laser phase noise, polarization tracking convergence error, quantization noise, and Carrier Phase Estimation (CPE) error~\cite{mecozzi2023use}.

\begin{figure*}[h!]
    \centering
    \includegraphics[width=0.84\linewidth, trim={2.1cm 0cm 2.1cm 0cm},
        clip]{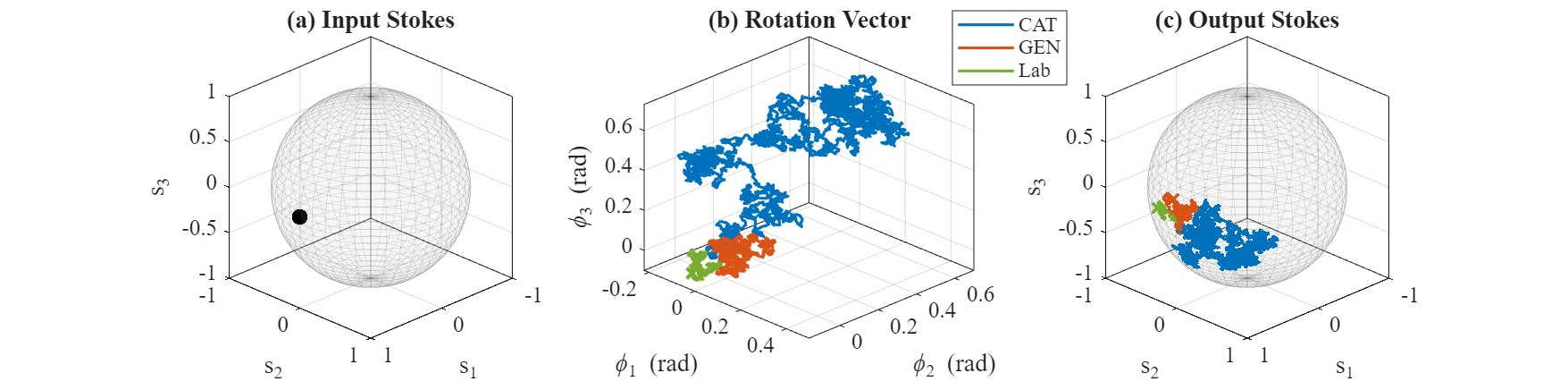}
    % \put(-245,195){(a)}
    % \put(-125,195){(b)}
    % \put(-245,95){(c)}
    % \put(-125,95){(d)}
    \caption{6-hours Stokes response to the modeled rotation vector using $J_k = J(\dot{\phi}_k) J_{k-1}$ with the variance inputs extracted from real datasets.}
    \label{fig:model}
\end{figure*}

Fig.~\ref{fig:psd} shows the evolution of the SOP on the Poincaré sphere for an arbitrary input state across all three datasets, i.e., (a) CAT, (b) GEN, and (c) Lab, over a 6-hour interval, together with the corresponding rotation vector total PSD computed over a 24-hour record in the second row of the plot. It can be observed that harmonic structures appear in the polarization PSD. These harmonics are clearly visible in the Lab dataset, while they are partially obscured in the field data due to the smoothing of PSD. We attribute these spectral features to strong transponder-related polarization oscillations, transformed by the nonlinear effects of polarization transformation in the large signal regime that generate multiple vanishing discrete tones spaced by a fundamental perturbation frequency. Also, the total PSD of rotation vectors is shown in the second row of Fig~\ref{fig:psd}. These results indicate that, after the possible removal of the white-noise floor, the detection sensitivity would be fundamentally limited by the random-walk polarization dynamics. Consequently, an improvement in sensitivity of up to 20~dB could be achieved at frequencies approaching the Nyquist limit of sampling.

Next, we evaluated the theoretical model introduced at the end of Section~\ref{sectionII} and originally proposed in Ref.~\cite{czegledi2016polarization}. Using the $\sigma_p^2$ mapping, we generated a pure random-walk process on SU(2). In this framework, a single parameter, $\sigma_p^2$, governs the stochastic evolution of polarization rotations in a three-dimensional space with three degrees of freedom. Mapping an input Stokes vector to an output Stokes vector through this rotation, however, inherently reduces the observable degrees of freedom by one. Figure~\ref{fig:model}(b) illustrates the 6-hour simulated SU(2) random-walk trajectories of rotation vector for the modeled CAT, GEN, and Lab transponders. It should be noted that these trajectories represent a polarization random walk on SU(2) and should not be confused with a Wiener process. For a given input Stokes vector, like the one shown in Fig.~\ref{fig:model}(a) , the corresponding evolution of the output Stokes vector under the generated transformations is shown in Fig.~\ref{fig:model}(c). It is important to distinguish between intrinsic polarization noise, arising from the native stochastic evolution of the optical field in the transmission, and extrinsic noise contributions that originate from other sources and are subsequently mapped into the SU(2) representation through the receiver processing chain. 

\section{Conclusion}
This work demonstrates that terminal-induced polarization noise is considerable for SOP sensing using coherent transponders. Analysis of field and laboratory data revealed polarization dynamics consisting of a low-frequency random walk and a higher-frequency noise floor, with noise levels increasing for longer links. The results highlight the need to identify and mitigate terminal noise sources to improve sensing sensitivity and further enable the use of existing submarine telecommunication cables for large-scale environmental and geophysical monitoring.

\bibliographystyle{plain} 
\bibliography{Refs}
\end{document}